\documentclass[10pt,conference]{IEEEtran}
\usepackage[utf8]{inputenc}
\usepackage{amsmath}
\usepackage{cite}
\usepackage{amssymb}
\usepackage{stackrel}
\usepackage{booktabs}
\usepackage{mathtools}
\usepackage{amsthm}
\usepackage{comment}
\usepackage{bm}
\usepackage{enumerate}
\usepackage{enumitem}
\usepackage[nolist]{acronym}
\usepackage{pgfplots}
\usepackage{microtype}
\usepackage{mathrsfs}
\usepackage{sansmath}
\usepackage{bbm}
\usepackage{amsfonts}
\pgfplotsset{compat=newest}
\pgfplotsset{plot coordinates/math parser=false}
\usepackage{tabularx}
\usepackage{float}
\usepackage{makecell}
\usepackage{collcell}
\usepackage{tikz}
\usepackage{setspace}
\usepackage{xcolor}
\usepackage{multirow}
\usepackage{graphicx}
\usetikzlibrary{decorations.pathreplacing,calc}
\usetikzlibrary{plotmarks}
\usetikzlibrary{matrix}
\usepackage[ruled,linesnumbered]{algorithm2e}
\usepackage{comment}
\usepackage{flushend}
\ifCLASSOPTIONcompsoc
    \usepackage[caption=false, font=normalsize, labelfont=sf, textfont=sf]{subfig}
\else
\usepackage[caption=false, font=footnotesize]{subfig}
\fi

\title{A Neural Network-aided Low Complexity \\ Chase Decoder for URLLC}

\author{
    \IEEEauthorblockN{Enrico~Testi and Enrico~Paolini}
    \IEEEauthorblockA{CNIT/WiLab, DEI, University of Bologna, Italy
    \\Email:\{enrico.testi, e.paolini\}@unibo.it}
}
\definecolor{carrotorange}{rgb}{0.93, 0.57, 0.13}

\DeclareMathOperator*{\argmin}{arg\,min}

\begin{acronym}
\acro{3GPP}{Third Generation Partnership Project}
\acro{ACK}{acknowledgment}
\acro{AGC}{asymmetric Granger causality}
\acro{AI}{artificial intelligence}
\acro{AIC}{Akaike information criterion}
\acro{AP}{access point}
\acro{AR}{auto-regressive}
\acro{AWGN}{additive white Gaussian noise}
\acro{BDD}{bounded distance decoder}
\acro{BPSK}{binary phase-shift keying}
\acro{BS}{base station}
\acro{BSS}{blind source separation}
\acro{CE}{conditional entropy}
\acro{CER}{codeword error rate}
\acro{CNN}{convolutional neural network}
\acro{CR}{coding rate}
\acro{CRC}{cyclic redundancy check}
\acro{CRLB}{Cramèr-Rao lower bound}
\acro{CSS}{Chirp Spread Spectrum}
\acro{CTE}{conditional transfer entropy}
\acro{DL}{deep learning}
\acro{DNN}{deep neural network}
\acro{DSP}{digital signal processor}
\acro{DtS}{direct-to-satellite}
\acro{ED}{end device}
\acro{EIRP}{effective isotropic radiated power}
\acro{EPMF}{empirical probability mass function}
\acro{FHSS}{frequency-hopping spread spectrum}
\acro{FIM}{Fisher's information matrix}
\acro{FLOPS}{floating point operations per second}
\acro{F-ICA}{fast independent component analysis}
\acro{GC}{Granger causality}
\acro{GDOP}{geometric diluition of precision}
\acro{GMSK}{Gaussian minimum-shift keying}
\acro{GPS}{Global Positioning System}
\acro{GPS-RTK}{Global Positioning System real-time kinematic}
\acro{GSP}{graph signal processing}
\acro{GW}{gateway}
\acro{HSMM}{hidden semi-Markov model}
\acro{KPCA}{kernel principal component analysis}
\acro{ICA}{independent component analysis}
\acro{IoT}{Internet of Things}
\acro{IoIT}{internet of intelligent things}
\acro{IoST}{Internet of Space Things}
\acro{ISM}{industrial, scientific and medical}
\acro{i.i.d.}{independent, identically distributed}
\acro{KL}{Kullback–Leibler}
\acro{LDPC}{low-density parity-check}
\acro{LDRO}{LowDataRateOptimization}
\acro{LEO}{low Earth orbit}
\acro{LOS}{line-of-sight}
\acro{LPWAN}{low-power wide-area network}
\acro{LRB}{least reliable bits}
\acro{LS}{least squares}
\acro{LoRa}{Long Range}
\acro{LoRaWAN}{Long Range Wide Area Network}
\acro{LR-FHSS}{Long Range Frequency Hopping Spread Spectrum}
\acro{MAC}{medium access protocol}
\acro{MC}{Monte Carlo}
\acro{MDL}{minimum description length}
\acro{MDS}{maximum distance separable}
\acro{ML}{maximum likelihood}
\acro{MLE}{maximum likelihood estimation}
\acro{MILP}{mixed integer linear programming}
\acro{NB-IoT}{Narrowband IoT}
\acro{NMS}{normalized min-sum}
\acro{NN}{neural network}
\acro{NLOS}{non-line-of-sight}
\acro{OBW}{occupied bandwidth}
\acro{NTN}{non-terrestrial network}
\acro{PCA}{principal component analysis}
\acro{p.d.f.}{probability density function}
\acro{PHY}{physical}
\acro{PPG}{perturbation pattern generation}
\acro{PPP}{Poisson point process}
\acro{RBF}{radial basis function}
\acro{ReLU}{rectified linear unit}
\acro{RF}{radio-frequency}
\acro{RL}{reinforcement learning}
\acro{RMSE}{root mean squared error}
\acro{RSS}{received signal strength}
\acro{RSSI}{received signal strength indicator}
\acro{r.v.}{random variable}
\acro{SF}{spreading factor}
\acro{SNR}{signal-to-noise ratio}
\acro{STDM}{statistical time-division multiplexing}
\acro{SU}{secondary user}
\acro{SVM}{support vector machine}
\acro{TE}{transfer entropy}
\acro{ToA}{time of arrival}
\acro{UAV}{unmanned aerial vehicle}
\acro{UE}{user equipment}
\acro{URLLC}{ultra-reliable low-latency communications}
\acro{UWB}{ultrawide band}
\end{acronym}  % include the list of acronyms from the file %%%%%%%%%%%%%%%%%%%%%%%%%%%%%%%%%%%%%%%%%%%%%%%%%%%%%%%%%%%%%

\begin{document}

\maketitle

\begin{abstract}
\Ac{URLLC} demand decoding algorithms that simultaneously offer high reliability and low complexity under stringent latency constraints. 
While iterative decoding schemes for LDPC and Polar codes offer a good compromise between performance and complexity, they fall short in approaching the theoretical performance limits in the typical \ac{URLLC} short block length regime.
Conversely, quasi-ML decoding schemes for algebraic codes, like Chase-II decoding, exhibit a smaller gap to optimum decoding but are computationally prohibitive for practical deployment in \ac{URLLC} systems. 
To bridge this gap, we propose an enhanced Chase-II decoding algorithm that leverages a \ac{NN} to predict promising perturbation patterns, drastically reducing the number of required decoding trials. 
The proposed approach combines the reliability of quasi-ML decoding with the efficiency of \ac{NN} inference, making it well-suited for time-sensitive and resource-constrained applications. 
\end{abstract}

\acresetall

\section{Introduction}

Emerging applications such as industrial automation and vehicular networks demand communication systems capable of achieving extremely high reliability with stringent latency requirements—a paradigm referred to as \ac{URLLC} \cite{KhaJanAhm:J22}. Designing efficient decoding algorithms for \ac{URLLC} remains a fundamental challenge, where reliability, latency, and %computational 
complexity must~be~balanced.

Iterative decoding algorithms for \ac{LDPC} and Polar codes, such as belief propagation and successive cancellation decoding, are widely adopted due to their favorable trade-off between performance and complexity \cite{YueMilShi:J23}, %These schemes iteratively refine decoding decisions, 
providing a practical solution for many \ac{URLLC} use cases. 
However, in the typical \ac{URLLC} short block length regime iterative decoders show visible losses with respect to theoretical limits, especially at low error rates.
Short algebraic codes—such as BCH and Reed-Solomon codes—paired with quasi-\ac{ML} decoding algorithms, tend to be very competitive in terms of error rate performance, although decoding complexity remains often incompatible with \ac{URLLC} requirements. 
Among such decoders, the Chase-II algorithm stands out, delivering near-\ac{ML} performance by exhaustively testing multiple perturbations of the hard-decision vector \cite{Chase:J72}. 
Despite its reliability advantages, the Chase-II decoder's exponential complexity makes it unsuitable for real-time deployment in latency-sensitive applications.

Recent advances in deep learning have opened new pathways for enhancing decoding algorithms by incorporating data-driven learning into traditional decoding frameworks. In \cite{NacMarLug:J18}, deep learning is shown to improve belief propagation and min-sum decoders by learning better message update rules. Recurrent neural architectures have also been proposed to reduce parameter complexity while retaining strong performance.
Several works have applied \acp{NN} to decoding of Polar codes. In \cite{LeeLeeFis:C22}, a \ac{DNN} is used to identify which bits to flip in belief propagation flip decoding. Similarly, \acp{CNN} have been explored to reduce decoding latency in  \cite{LiTiaZha:C22}. Other approaches leverage path metric information to simplify successive cancellation list decoding, yielding near-optimal performance with reduced complexity \cite{LuZhaLei:J23}.

In the context of algebraic codes, learning-based decoding has also shown promise. The work in \cite{BenChoKis:C18} proposes a \ac{DNN} that uses syndromes and soft reliabilities to estimate channel noise, improving generalization and reducing overfitting. A different perspective is taken in \cite{ZhuCaoZha:C20}, where a \ac{DNN} is trained as a denoiser to directly map noisy codewords to their clean counterparts.

Building on this, our work focuses on reducing the complexity of quasi-\ac{ML} decoding—specifically, the Chase-II algorithm—by leveraging machine learning. We propose a \ac{NN}-aided Chase-II decoder that significantly reduces the number of required perturbation trials by learning to predict the most likely error patterns. The decoder maintains the high reliability of the Chase-II algorithm while achieving substantial computational savings.

The main contributions of this paper are summarized as follows:
\begin{itemize}
    \item We introduce a novel \ac{NN}-aided Chase-II decoding framework that achieves near-optimal decoding performance with significantly reduced complexity, making it well suited for \ac{URLLC} scenarios.
    
    \item We develop and train a \ac{NN} that learns to predict the most promising perturbation patterns, enabling a selective refinement strategy (named \ac{NN}-$\rho$) to limit the number of decoding trials.
    
    \item We provide a comprehensive complexity analysis in terms of \ac{FLOPS} and compare the runtime performance of the proposed scheme with traditional Chase-II and \ac{BDD}, showing substantial execution time savings.
    
    %\item Through simulations over a BCH$(127,64)$ code, we demonstrate that the proposed decoder matches the performance of Chase-II with up to $85\%$ lower complexity.
\end{itemize}
%%%%%%%%%%% DRAFT %%%%%%%%%%%%%

\section{System Model}\label{sec:system_model}
The user aims to transmit a $k$-bit information word $\mathbf{u} \in \{0,1\}^k$ using a binary linear block code $\mathcal{C}$ of length $n$, dimension $k$, and minimum distance $d_\text{min}$. Let $\mathbf{G}$ denote the generator matrix of code $\mathcal{C}$. The information word $\mathbf{u}$ is encoded into the codeword $\mathbf{c} = \mathbf{u} \mathbf{G} \in \{0,1\}^n$. Then, the encoded sequence is modulated using \ac{BPSK} modulation through the mapping $\tau : \{0,1\}\mapsto\{+1,-1\}$, resulting in the transmitted modulated codeword 
\begin{align} 
\mathbf{x} = \tau(\mathbf{c}). 
\end{align} 
The modulated codeword is transmitted over \ac{AWGN} channel, such that the received soft-symbols $\mathbf{y} \in \mathbb{R}^n$ can be expressed as
\begin{align}
    \mathbf{y} = \mathbf{x} + \mathbf{w}
\end{align}
where $\mathbf{w}\sim \mathcal{N}(\mathbf{0},\sigma_\text{N}^2 \mathbf{I}_n)$ is the %noise vector. 
vector of noise samples.
The log-likelihood ratios of the received symbols, $\boldsymbol{\gamma}=(\gamma_1,\dots,\gamma_n)$, can be expressed as
\begin{align}
 \gamma_i = \log{\frac{p(y_i|c_i=0)}{p(y_i|c_i=1)}}=\frac{2y_i}{\sigma_\text{N}^2} \, , \,\,\, i=1,\dots,n.   
\end{align}
%
%where $p(z)$ is the probability density function of a \ac{r.v.} distributed as $\mathcal{N}(0,\sigma_\text{N}^2)$. 
Let $\mathbf{r}=(|\gamma_1|,|\gamma_2|,\dots,|\gamma_n|)$ be the vector of reliabilities corresponding to the received codeword symbols, where $|\gamma_i|$ indicates the reliability of the $i$th symbol. The vector of hard decision symbols, $\mathbf{q} \in \{0,1\}^n$, is obtained by applying a hard decision to the received sequence $\mathbf{y}$. Specifically, $q_i = 0$ if $y_i\geq0$, and $q_i = 1$ otherwise. We also let $\mathbf{e}=\mathbf{c} \oplus \mathbf{q}$ be the error pattern derived after performing binary decoding on $\mathbf{q}$.

\subsection{Chase-II Decoder Overview}\label{sec:Chase-II}
The Chase-II algorithm stands out among algebraic soft-decision decoding methods designed to improve the decoding performance of linear block codes, such as BCH 
%and Reed-Solomon 
codes. 
The Chase-II decoding algorithm can be summarized as follows:
\begin{itemize}
    \item The decoding process starts by analyzing the received soft information to identify the \acp{LRB}, i.e., the bits with the highest uncertainty. Those are the $t=\lfloor\frac{d_\text{min}-1}{2}\rfloor$ bits having the lowest reliabilities. 
    The received soft symbols are sorted in ascending order of magnitude, resulting in a permutation $\pi(\mathbf{y})$. 
    The same operation is performed to the reliabilities and the hard decision vectors, yielding $\pi(\boldsymbol{\gamma})$, and $\pi(\mathbf{q})$. 
    The first $t$ entries of the ordered vectors correspond to the \acp{LRB}.
    
    \item Once the \acp{LRB} are identified, the Chase-II decoder systematically generates and tests all possible perturbation patterns for these bits. Each perturbation pattern corresponds to a hypothesis about the transmitted codeword, and the decoder evaluates them to determine the most likely candidate. 
    We denote by $\mathcal{P}$ the set %containing 
    of all %the 
    possible perturbation patterns $\mathbf{p}_j \in \{0,1\}^{n}$ with $j=1,\dots,2^{t}$, i.e., sequences that contain $1$'s in the location of the \acp{LRB} %that have 
    to be inverted. 
    By adding a test pattern, modulo-$2$, to the received hard symbols, we obtain a new sequence
    \begin{align}
        \tilde{\mathbf{q}}_j = \mathbf{q} \oplus \mathbf{p}_j
    \end{align}
    and by running a \ac{BDD} a new error pattern %$\tilde{\mathbf{e}}_j$ 
    is obtained. 
    Such an error pattern may or may not %be different 
    differ from the original error pattern $\mathbf{e}$ depending on wether or not $\tilde{\mathbf{q}}_j$ falls within the decoding sphere of a new codeword.

    \item When the $j$th test pattern produces a valid codeword $\hat{\mathbf{c}}_j$ as the output of the \ac{BDD}, the distance between this codeword and the received soft symbols $\mathbf{y}$ is computed. Specifically, for the $j$th pattern, the weighted distance is calculated as $w_j=|\mathbf{y}-\tau(\hat{\mathbf{c}}_j)|_1$, where $|\cdot|_1$ denotes the $\ell_1$ norm of the vector. This process is repeated for all the generated test patterns, and the pattern that results in the smallest weighted distance is selected as
\begin{align}
 l = \argmin_{j} w_j.
\end{align}
    The corresponding estimated codeword, $\hat{\mathbf{c}}_l$ is then provided as the output of the decoder.
\end{itemize}

%The exhaustive search approach enables the Chase-II decoder to achieve near-\ac{ML} performance, making it a strong choice for scenarios demanding high reliability.
%However, the computational complexity of this approach grows exponentially with the number of \ac{LRB}. For every additional \ac{LRB}, the number of perturbation patterns doubles, imposing a significant computational burden. This limitation makes the direct application of the Chase-II decoder impractical for ultra-low latency scenarios, such as \ac{URLLC}, where almost real-time decoding is critical.

The exhaustive search approach enables the Chase-II decoder to achieve near-\ac{ML} performance, making it suitable for high-reliability applications. However, evaluating all possible $2^t$ perturbation patterns imposes substantial computational complexity, which grows exponentially with the number of \acp{LRB}. This complexity makes Chase-II decoding impractical for ultra-low latency scenarios, such as \ac{URLLC}, where real-time decoding is essential.

\begin{figure*}[th]
    \centering
    \includegraphics[width=0.8\linewidth]{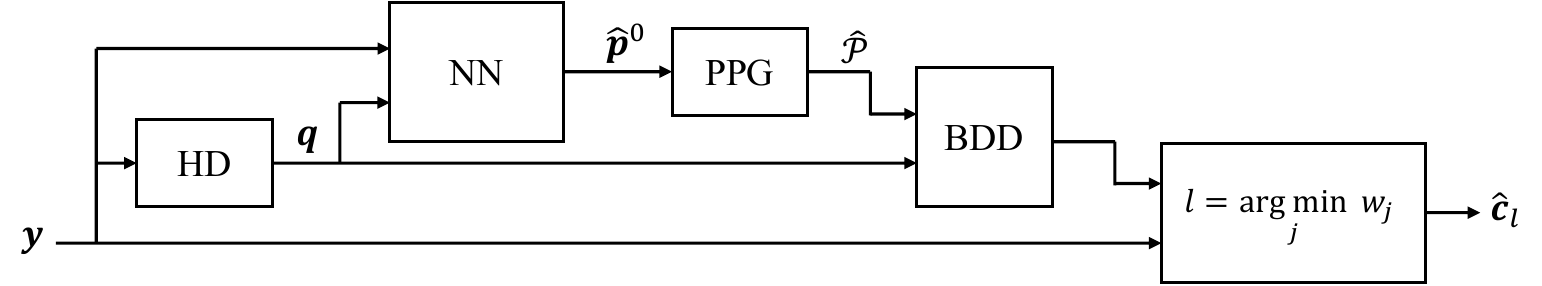}
    \caption{Logical block diagram of the proposed NN-aided Chase-II decoder. The received soft symbols, the hard-decision vector, and reliability information are fed into the neural network, which predicts a base perturbation pattern over the LRBs. The PPG block expands this base pattern by systematically flipping small subsets of LRBs, producing a refined set of candidate test patterns. Each pattern is combined with the hard-decision vector and decoded via a BDD. Among the decoded codewords, the one with minimum soft distance to the received symbols is selected as the final output.}
    \label{fig:NChase_scheme}
\end{figure*}
\section{Neural Network-aided Chase-II Decoding}\label{sec:DLChase}
To reduce the computational complexity of the traditional Chase-II decoder while preserving near-optimal decoding performance, we propose a novel algorithm that integrates an \ac{NN} into the decoding pipeline. The core idea is to use the \ac{NN} to predict the most promising perturbation pattern, thereby reducing the number of test patterns that must be evaluated during decoding. This results in substantial computational savings, making the approach particularly attractive for latency- and resource-constrained applications such as \ac{URLLC}.

\subsection{Overview of the NN-Aided Decoding Framework}
The proposed decoding algorithm leverages a trained \ac{NN} to guide the perturbation pattern selection process. 
During training, the network learns to mimic the decisions of %a classical 
an actual Chase-II decoder by observing perturbation patterns that result in correct codeword recovery. Once trained, the \ac{NN} acts as a lightweight yet effective approximation of the Chase-II exhaustive search, providing a  perturbation pattern that is likely to lead to successful decoding.

At inference time, given the received soft symbol vector $\mathbf{y}$, the decoder first identifies the set of \acp{LRB}. The neural network processes a structured representation of the received data—including soft symbols, reliabilities, and hard decisions—and outputs a base perturbation pattern $\hat{\mathbf{p}}^0 \in \mathcal{P}$. This pattern serves as the central candidate around which the decoder builds a refined set of test patterns.
Specifically, starting from the \ac{NN}-predicted base pattern, an expanded set of candidate perturbation patterns is generated by systematically flipping all subsets of \acp{LRB} of size up to $\rho$. We refer to this algorithm as \ac{NN}-$\rho$, which
%To enhance robustness, additional candidate patterns are generated by flipping subsets of the \acp{LRB}. We define the algorithm \ac{NN}-$\rho$ as the scheme that evaluates all patterns formed by flipping up to \( \rho \) bits in the \ac{NN}-predicted pattern. This 
results in a total number of test patterns given by
%The \ac{NN} is trained to replicate the decision-making process of the Chase-II decoder. Specifically, during the training phase, the NN learns to predict the most promising perturbation pattern by observing perturbations generated by exhaustive search. Once trained, the \ac{NN} provides a lightweight yet effective approximation of the exhaustive search, drastically reducing the number of patterns that must be explicitly tested by the decoder.
%At decoding time, given the received symbols $\mathbf{y}$, the decoder first identifies the \acp{LRB}. The neural network then processes the received soft symbols, the reliabilities, and the hard decision vector and outputs a base perturbation pattern, $\hat{\mathbf{p}}^0$, that is most likely to yield the correct codeword. This predicted pattern serves as the central candidate from which additional patterns can be generated by individually flipping tuples of bit in the set of the identified \acp{LRB}. 
%Specifically, starting from the \ac{NN}-predicted base pattern, an expanded set of candidate perturbation patterns is generated by systematically flipping subsets of \acp{LRB} of size up to $\rho$. We refer to this approach as NN-$\rho$, where the algorithm, after applying the \ac{NN}-predicted pattern, generates candidate test patterns by flipping all possible tuples of $1, 2, \dots, \rho$ bits among the identified \acp{LRB}. The total number of test patterns evaluated is:
%
\begin{equation}
    N_\text{T} = \sum_{i=0}^{\rho}\binom{t}{i}. 
\end{equation}
Each test pattern is combined modulo-2 with the hard-decision vector $\mathbf{q}$, and the resulting sequence is decoded using a \ac{BDD}. 
Among all decoded codewords, the one with the minimum soft-distance to $\mathbf{y}$ is selected as the final output.
The decoding flow is illustrated in the block diagram in Fig.~\ref{fig:NChase_scheme}, where the \ac{PPG} module produces the candidate set from the \ac{NN} base prediction. The complete procedure is also detailed in the pseudo-code in Algorithm~\ref{alg:NN-ChaseII}.

Compared to the exhaustive Chase-II algorithm, which tests all $2^t$ patterns, the \ac{NN}-$\rho$ scheme drastically narrows the search space while preserving decoding performance. The \ac{NN}-aided decoder thus achieves a favorable trade-off between complexity and reliability, making it well suited for scenarios requiring fast and efficient decoding.

%These patterns are combined, modulo-$2$, with the hard-decision vector derived from $\mathbf{y}$ to produce candidate vectors. Each candidate vector undergoes decoding via \ac{BDD}, and the corresponding codewords (if any) are evaluated based on their distances from the received soft symbols. The codeword resulting in the smallest distance is selected as the final decoded codeword. The block diagram illustrating the detailed functioning of the proposed \ac{NN}-aided Chase-II decoding scheme is presented in Fig.~\ref{fig:NChase_scheme}, where the generation of the test patterns via perturbation of the \ac{NN} output is denoted as \ac{PPG}. The pseudo-code of \ac{NN}-$\rho$ is given in Algorithm~\ref{alg:NN-ChaseII}.
%Compared to the exhaustive search in the traditional Chase-II decoder, this \ac{NN}-aided approach significantly reduces the computational burden by drastically narrowing the perturbation search space. Rather than testing every possible perturbation pattern, the algorithm tests only the \ac{NN}-predicted pattern and the ones obtained by perturbing few tuples of bits of the \ac{NN} output, effectively combining low complexity with near-optimal performance suitable for \ac{URLLC}.
%The \ac{NN} structure, as well as its input and output details, are detailed in Section~\ref{sec:NN_architecture}, while the computational complexity analysis of the proposed decoder is discussed in detail in Section~\ref{sec:complexity}.

\begin{algorithm}[t]
\caption{Pseudo-code for the NN-aided Chase-II Decoder}\label{alg:NN-ChaseII}
%\small
\DontPrintSemicolon
\KwIn{Received signal $\mathbf{y}$, reliability vector $\boldsymbol{\gamma}$, hard decision vector $\mathbf{q}$, $\rho$}
\KwOut{Estimated codeword $\hat{\mathbf{c}}_{l}$}
%Identify the $N_T$ LRBs of $\mathbf{q}$\;
$\boldsymbol{\Omega} \gets (\mathbf{y};\boldsymbol{\gamma}; \mathbf{q};\pi(\mathbf{y});\pi(\boldsymbol{\gamma});\pi(\mathbf{q}))$\;
Generate base test pattern $\hat{\mathbf{p}}^{0} \gets \text{NN}(\boldsymbol{\Omega})$\;
Generate the set of test patterns $\hat{\mathcal{P}}$, with $|\hat{\mathcal{P}}|=N_\text{T}$, by flipping all the subsets of bits of $\hat{\mathbf{p}}^{0}$ of size up to $\rho$\;
\ForEach{$\hat{\mathbf{p}}_j \in \hat{\mathcal{P}}$}{
    Generate perturbed vector $\tilde{\mathbf{q}}_j \gets \mathbf{q} \oplus \hat{\mathbf{p}}_j$\;
    Perform BDD on $\tilde{\mathbf{q}}_j$\;
    \If{decoding successful}{
        Obtain codeword candidate $\hat{\mathbf{c}}_j$\;
        Compute the metric $w_j \gets |\mathbf{y}-\hat{\mathbf{c}}_j|_1$\;
    }
}
$l \gets \argmin_{j} w_j$\;
\KwRet{$\hat{\mathbf{c}}_{l}$}\;
\label{alg:NN-Chase-II}
\end{algorithm}

\subsection{NN Architecture} \label{sec:NN_architecture}
The perturbation prediction task is cast as a multi-label classification problem, where each of the $t$ bits in the perturbation pattern is independently predicted. This formulation allows the network to selectively identify likely error positions within the set of \acp{LRB}, replacing the need for a full combinatorial search.

The network’s input is the  feature vector
\begin{equation}
\boldsymbol{\Omega} = \left(\mathbf{y}, \boldsymbol{\gamma}, \mathbf{q}, \pi(\mathbf{y}), \pi(\boldsymbol{\gamma}), \pi(\mathbf{q})\right) \in \mathbb{R}^{6n}
\end{equation}
which includes soft symbols $\mathbf{y}$, their reliability $ \boldsymbol{\gamma}$, and the hard decision vector $\mathbf{q}$, along with their permutations under the sorting map $\pi(\cdot)$ that ranks bits by reliability. 

The network consists of a single fully connected hidden layer of size $N_\text{H}$, followed by a \ac{ReLU} activation. The output layer produces $t$ soft values using sigmoid activation
\begin{align}
\sigma(z) = \frac{1}{1 + e^{-z}}
\end{align}
indicating the likelihood of flipping each bit in the perturbation pattern. The network is trained using binary cross-entropy loss
\begin{align}
\mathcal{L}(\mathbf{p}, \hat{\mathbf{p}}^0) = -\sum_{i=1}^{t} \left( p_i \log(\hat{p}^{0}_{i}) + (1 - p_i) \log(1 - \hat{p}^{0}_{i}) \right)
\end{align}
where $\mathbf{p}$ is the true perturbation pattern obtained from the Chase-II decoder and $\hat{\mathbf{p}}^0$ is the base one.
After thresholding the sigmoid outputs at $0.5$, a binary perturbation vector is obtained. This predicted pattern is zero-padded (on the non-\ac{LRB} positions) and reordered via inverse permutation, $\pi^{-1}(\cdot)$, to match the original bit positions. The resulting binary mask is then applied to the hard-decision vector during decoding.
The neural network structure is summarized in Table~\ref{tab:architecture}.

\begin{table}[t]
   \caption{NN architecture}
  \label{tab:architecture}
  \centering
  \normalsize
    \begin{tabular}{l|l|l}
    \toprule
    \textbf{Layer type} & \textbf{Layer size} & \textbf{Activation}\\
    \midrule
    Input & $6n$ &  \\
    Fully connected & $N_\text{H}$ & ReLu \\
    Output & $t$ & Sigmoid \\
    
    \bottomrule
 \end{tabular}
\end{table}

\subsection{Decoding Complexity} \label{sec:complexity}

\subsubsection{Bounded Distance Decoder (BDD)}
The computational complexity of a \ac{BDD} depends on the specific decoding algorithm used and can be measured in terms of finite-field operations, typically multiplications and additions. Decoding BCH codes via bounded distance decoding involves four main stages: syndrome computation, solving the Berlekamp–Massey algorithm, running Chien search, and performing error correction using Forney algorithm.
The syndrome computation requires evaluating $2t$ syndromes, each involving approximately $n$ multiplications and $n-1$ additions \cite{SchEliRos:C11}. Thus, this stage has a complexity of $2t(2n-1)$ \ac{FLOPS}.
Solving the Berlekamp–Massey algorithm, which yields the error locator polynomial, has a computational complexity of approximately $t^2$ multiplications and $t^2$ additions \cite{ImaYos:J87}. Finding the roots of the error locator polinomial via Chien method requires $nt$ additions and $nt$ multiplications \cite{Chien:J64}. Finally, Forney’s algorithm, used for actual error correction, also involves approximately $t^2$ multiplications and $t^2$ additions \cite{Forney:J65}.
Therefore, the total computational complexity of the \ac{BDD} decoder is approximately 
\begin{align}
    C_{\mathrm{BDD}} = 6nt+2t(2t-1)
\end{align} \ac{FLOPS}.

\subsubsection{Chase-II decoder}
Considering that the Chase-II algorithm involves running $2^{t}$ \acp{BDD}, and evaluating the weighted distance for each codeword obtained, its overall complexity is 
\begin{align}
    C_{\mathrm{Chase}} = 2^{t} C_{\mathrm{BDD}}
\end{align}
\ac{FLOPS}.

\subsubsection{NN-aided Chase-II decoder}
The computational complexity of the \ac{NN} in terms of \ac{FLOPS} is computed as follows. The fully connected hidden layer requires $12 N_\text{H} n$ \ac{FLOPS} due to the matrix multiplication, while the subsequent ReLU activation contributes additional $N_\text{H}$ \ac{FLOPS}. Similarly, the fully connected output layer has complexity $2 t N_\text{H}$ \ac{FLOPS}, followed by a sigmoid activation adding $4 t$ \ac{FLOPS}. Therefore, the total complexity for the \ac{NN} is
\begin{align}
   C_{\mathrm{NN}} =  12 N_\text{H} n + 2 t N_\text{H} + N_\text{H} + 4 t
\end{align}
\ac{FLOPS}.
Considering that the \ac{NN}-aided Chase-II decoder executes the \ac{BDD} a total of $N_\text{T}$ times and selects the codeword of minimum weight, the overall complexity is
\begin{equation} \label{eq:complexity_CPU}
    C_{\mathrm{NN-}\rho} = N_\text{T}\, C_{\mathrm{BDD}} + C_{\mathrm{NN}}
\end{equation}
\ac{FLOPS}. Considering that \ac{NN} inference can be efficiently parallelized on hardware platforms equipped with dedicated neural processing units, such as GPUs or specialized accelerators, the dominant component in \eqref{eq:complexity_CPU} is the one related to the execution of the \ac{BDD}, so that the total complexity can be further simplified as
\begin{equation} \label{eq:complexity_GPU}
    C_{\mathrm{NN-}\rho} \approx N_\text{T} \, C_{\mathrm{BDD}}
\end{equation}
FLOPS.

\section{Numerical Results}\label{sec:numres}
This section presents a performance evaluation of the proposed \ac{NN}-aided Chase-II decoding scheme introduced in Section~\ref{sec:DLChase}. The performance is assessed through extensive Monte Carlo simulations and compared with the classical Chase-II decoder and the standard \ac{BDD} in terms of decoding reliability, computational complexity, and execution time.

\subsection{Simulation Setup}
The simulations are conducted on a binary BCH code with $n=127$, $k=64$ and minimum distance $d_\text{min} = 21$, which allows correction of up to $t = 10$ errors. A distinct neural network is trained and used for each simulated \ac{SNR} value. For each \ac{SNR}, the Chase-II decoder is employed to generate training and test datasets composed of $1.6 \times 10^6$ and $10 \times 10^6$ samples, respectively.
Each network is initialized using He initialization \cite{HeZhaRen:15a} and trained using the Adam optimizer \cite{KinBa:15} over $130$ epochs with mini-batches of $1024$ samples. The initial learning rate is set to $10^{-3}$ and reduced by a factor of $10$ every $50$ epochs. After empirical tuning, the number of neurons in the hidden layer is fixed to $N_\text{H} = 512$ to balance performance and complexity.

\begin{figure}
    \centering
    \resizebox{0.99\columnwidth}{!}{\definecolor{aloha1}{rgb}{0.89, 0.82, 0.04}
\definecolor{baseline}{rgb}{0.00000,0.44700,0.74100}%
\definecolor{genie}{rgb}{0.85000,0.32500,0.09800}%
\definecolor{radius30}{rgb}{0.59, 0.29, 0.0}%
\definecolor{radius40}{rgb}{0.0, 0.5, 0.0}%
\definecolor{radius80}{rgb}{0.0, 0.0, 0.0}%
\definecolor{corr2}{rgb}{1.0, 0.49, 0.0}%
\definecolor{corr1}{rgb}{0.6, 0.4, 0.8}

\begin{tikzpicture}
 
\begin{axis} [
    %xlabel style={font=\tiny},
    %tick label style={font=\tiny},
    %height = .5\linewidth,
    %width = \linewidth,
    axis on top,
    ybar = .1cm,
    bar width = 5pt,
    %ymode=log,
    %log basis y={10},
    ymin = 0, 
    ymax = 1, 
    ytick = {0,0.2,...,1},
    minor ytick = {0,0.1,...,1},
    ymajorgrids = true,
    yminorgrids = true,
    major grid style={solid, line width=0.1pt},
    minor grid style={dotted, line width=0.1pt, gray},
    axis on top = false,
    ylabel={\small EPMF},
    xlabel near ticks,
    xlabel style={font=\footnotesize},
    ylabel near ticks,
    xtick pos=bottom,
    enlarge x limits=0.1,
    xtick=data,
    symbolic x coords={{\footnotesize $d_\text{H}=1$}, {\footnotesize $d_\text{H}=2$}, {\footnotesize $d_\text{H}=3$},{\footnotesize $d_\text{H}=4$}, {\footnotesize $d_\text{H}=5$}},
    legend image post style={scale=0.4},
    legend columns=1,
    legend style={font=\normalsize,at={(.99,0.990)},nodes={scale=0.7, transform shape}},
    legend cell align={left},
]

\addplot[color=radius40, fill=radius40] coordinates {({\footnotesize $d_\text{H}=1$}, 0.638039) ({\footnotesize $d_\text{H}=2$},0.215719) ({\footnotesize $d_\text{H}=3$},0.0881339) ({\footnotesize $d_\text{H}=4$},0.0381875) ({\footnotesize $d_\text{H}=5$},0.0166616)};

\addplot[color=genie, fill=genie] coordinates {({\footnotesize $d_\text{H}=1$},0.794118) ({\footnotesize $d_\text{H}=2$},0.0784314) ({\footnotesize $d_\text{H}=3$},0.0784314) ({\footnotesize $d_\text{H}=4$},0.0245098) ({\footnotesize $d_\text{H}=5$},0.0245098)};

\addplot[color=baseline, fill=baseline] coordinates {({\footnotesize $d_\text{H}=1$}, 0.751992) ({\footnotesize $d_\text{H}=2$},0.179526) ({\footnotesize $d_\text{H}=3$},0.0623298) ({\footnotesize $d_\text{H}=4$},0.00619229) ({\footnotesize $d_\text{H}=5$},0)};

%\addplot[color=radius40, fill=radius40] coordinates {({\footnotesize $d_\text{H}=1$}, 0) ({\footnotesize $d_\text{H}=2$},0) ({\footnotesize $d_\text{H}=3$},0) ({\footnotesize $d_\text{H}=4$},0) ({\footnotesize $d_\text{H}=5$},0)};

%\addplot[color=baseline, fill=baseline] coordinates {($\sigma=3\,$dB,0.9967) ($\sigma=8\,$dB,0.9656)};

%\legend{$\text{SNR}=3\,$dB,$\text{SNR}=4\,$dB,$\text{SNR}=5\,$dB}
\legend{$\text{SNR}=2\,$dB,$\text{SNR}=3\,$dB,$\text{SNR}=4\,$dB}
\end{axis}
\end{tikzpicture}}
    \caption{EPMF
    of the Hamming distance between the \ac{NN}-predicted test patterns and the ones selected by the Chase-II decoder, conditioned on prediction errors, for various \ac{SNR} values.}
    \label{fig:EPMF}
\end{figure}
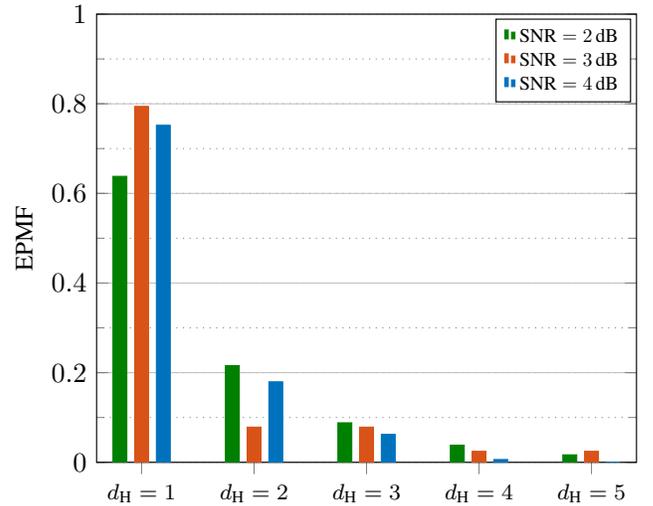

\subsection{Decoding Performance}

\subsubsection{Neural network behavior and pattern refinement}
To evaluate the quality of the \ac{NN} predictions, we analyze the distribution of Hamming distances between the predicted test patterns 
%(i.e., $\hat{\mathbf{p}}^0$) 
and the corresponding ones selected by the Chase-II decoder. Fig.~\ref{fig:EPMF} reports the \ac{EPMF} of these distances, conditioned on the event that the \ac{NN} fails to predict the correct test pattern.
The results show that the majority of mispredicted patterns are only one bit flip away from the correct one (i.e., $d_\text{H} = 1$), across all \ac{SNR} regimes. This supports the inclusion of the \ac{PPG} module, which systematically flips combinations of the least reliable bits to form additional candidate test patterns. Specifically, running the \ac{NN}-$1$ algorithm—where all single-bit flips of the \ac{NN} prediction are tested—captures most of the correct patterns while requiring a significantly smaller candidate set than the exhaustive Chase-II.
Furthermore, the probability of mispredicting more than four bits vanishes at high \ac{SNR}, which is attributed to the sparsity of the true error pattern in these regimes. This justifies limiting the perturbation depth to small values of $\rho$ in the NN-$\rho$ decoding framework.

\subsubsection{Decoding performance comparison}
\begin{figure}
    \centering
    \resizebox{0.99\columnwidth}{!}{
    \definecolor{aloha1}{rgb}{0.89, 0.82, 0.04}
\definecolor{baseline}{rgb}{0.00000,0.44700,0.74100}%
\definecolor{genie}{rgb}{0.85000,0.32500,0.09800}%
\definecolor{radius30}{rgb}{0.59, 0.29, 0.0}%
\definecolor{radius40}{rgb}{0.0, 0.5, 0.0}%
\definecolor{radius80}{rgb}{0.0, 0.0, 0.0}%
\definecolor{corr2}{rgb}{1.0, 0.49, 0.0}%
\definecolor{corr1}{rgb}{0.6, 0.4, 0.8}
%\definecolor{amber}
%
\begin{tikzpicture}

\begin{axis}[%
scale only axis,
xmin=1,
xmax=6,
xtick distance = 1,
xlabel style={font=\large},
xlabel={SNR [dB]},
%xlabel shift=-5pt,
ticklabel style = {font=\large},
ymode=log,
ymin=1e-5,
ymax=1,
yminorticks=true,
ylabel style={font=\large},
ylabel={CER},
axis background/.style={fill=white},
xmajorgrids,
ymajorgrids,
yminorgrids,
legend style={at={(0.01,0.41)}, anchor=north west, legend cell align=left, align=left, draw=white!15!black, fill opacity=0.8},
legend entries={%
%SIC, % linea continua
%no SIC, % linea tratteggiata
{BCH$(127,64)$, Chase-II},
{BCH$(127,64)$, NN-$0$},
{BCH$(127,64)$, NN-$1$},
{BCH$(127,64)$, NN-$2$},
{BCH$(127,64)$, BDD},
{LDPC$(128,64)$, NMS},
}
]

% Linea continua per SIC
%\addlegendimage{line legend, color=gray, line width=1.6pt}
% Linea tratteggiata per no SIC
%\addlegendimage{dashed, color=gray, line width=1.6pt}

% Marker per i valori di R
\addlegendimage{only marks, line width=1.4pt, mark=o, color=radius80, mark size=3pt,  mark options={solid, radius80}}
\addlegendimage{only marks, line width=1.4pt, mark=x, color=radius40, mark size=3pt, mark options={solid, radius40}}
\addlegendimage{only marks, line width=1.4pt, mark=x, color=corr1, mark size=3pt, mark options={solid, corr1}}
\addlegendimage{only marks, line width=1.4pt, mark=x, color=genie, mark size=3pt, mark options={solid, genie}}
\addlegendimage{only marks, line width=1.4pt, mark=diamond, color=baseline, mark size=3.2pt, mark options={solid, baseline}}
\addlegendimage{only marks, line width=1.4pt, mark=triangle, color=corr2, mark size=3pt, mark options={solid, corr2}}
    
%\addlegendentry{$K=25$, no SIC}
%Chase-II
\addplot [color=radius80, line width=1.8pt, mark size=2.7pt, mark=o, mark options={solid, radius80}, clip mode=individual]
  table[row sep=crcr]{%
1 0.6830\\
2 0.3080\\
3 0.0480\\
4 0.0011\\
5 1e-5\\
};

\addplot [color=baseline, line width=1.8pt, mark size=3pt, mark=diamond, mark options={solid, baseline}, clip mode=individual]
  table[row sep=crcr]{%
1 0.9440\\
2 0.8000\\
3 0.4300\\
4 0.0920\\
5 0.0100\\
6 0.0002\\
};

\addplot [color=corr2, line width=1.8pt, mark size=3.2pt, mark=triangle, mark options={solid, corr2}, clip mode=individual]
  table[row sep=crcr]{%
1 0.78\\
2 0.400\\
3 0.07200\\
4 0.003800\\
5 0.000040\\
};

%%%%% NChase
\addplot [color=radius40, line width=1.8pt, mark size=3.2pt, mark=x, mark options={solid, radius40}, clip mode=individual]
  table[row sep=crcr]{%
1 0.7330\\
2 0.4280\\
3 0.0792\\
4 0.0046\\
5 0.000109\\
};

%% NChase 1 bit flip
\addplot [color=corr1, line width=1.8pt, mark size=3.2pt, mark=x, mark options={solid, corr1}, clip mode=individual]
  table[row sep=crcr]{%
1 0.6930\\
2 0.3280\\
3 0.048800\\
4 0.0012\\
5 0.000019\\
};

%% NChase 2 bit flip
\addplot [color=genie, line width=1.8pt, mark size=3.2pt, mark=x, mark options={solid, genie}, clip mode=individual]
  table[row sep=crcr]{%
1 0.6880\\
2 0.3100\\
3 0.04840\\
4 0.0011\\
5 0.000013\\
};

\end{axis}
\end{tikzpicture}%
    }
    \caption{\ac{CER} varying the \ac{SNR} for the BCH$(127,64)$ code using the proposed NN-aided Chase-II decoders (NN-$0$, NN-$1$, NN-$2$), the classical Chase-II algorithm, and the baseline \ac{BDD}. The figure also includes a performance curve for the LDPC$(128,64)$ code from the CCSDS standard, decoded using the NMS algorithm.}
    \label{fig:performance_comparison}
\end{figure}
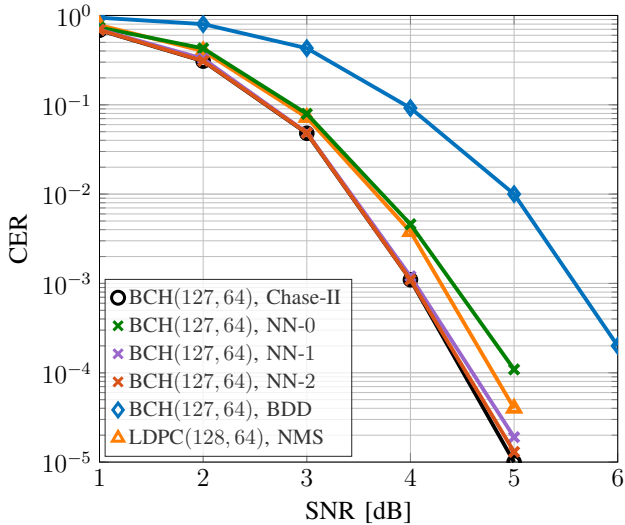

Fig.~\ref{fig:performance_comparison} presents the \ac{CER} as a function of the \ac{SNR} for several decoding algorithms. The NN-$1$ and NN-$2$ decoders achieve nearly identical performance to the full Chase-II decoder, with NN-$1$ exhibiting a slight performance degradation in the high-\ac{SNR} regime. Both schemes outperform \ac{BDD} by a substantial margin, achieving over $1.5\,$dB of \ac{SNR} gain at \ac{CER} $10^{-4}$.
While the NN-$0$ variant, which relies solely on the base prediction of the neural network, does not match the Chase-II performance, it still offers a notable improvement over the standalone \ac{BDD}, confirming the added value of the learned base pattern.
Additionally, the figure includes a performance curve for the LDPC$(128,64)$ code from the CCSDS standard \cite{CCSDS231.0-B-3-S}, decoded using the \ac{NMS} algorithm with a normalization factor of $0.75$ and a maximum of $100$ iterations. Interestingly, the NN-$0$ decoder for the BCH code achieves a performance comparable to that of the LDPC code under these conditions, despite being a one-shot decoder without iterative processing. This result highlights the efficiency of the proposed architecture. Furthermore, the BCH code with both NN-$1$ and NN-$2$ outperforms the LDPC one with \ac{NMS}, demonstrating the strong potential of the proposed neural decoding strategies even when compared with well-established iterative decoders.

\subsection{Computational Complexity and Runtime}
The computational complexity of each decoder is evaluated in terms of the number of \ac{FLOPS}, based on the analytical expressions provided in Section~\ref{sec:complexity}. In particular, the complexity of the NN-$\rho$ algorithms is assessed using~\eqref{eq:complexity_CPU}, which reflects execution on a general-purpose CPU. Table~\ref{tab:complexity} summarizes the total number of \ac{FLOPS} required for each decoder in the considered case study. As expected, the Chase-II decoder exhibits the highest complexity due to the exponential growth in the number of test patterns with respect to the number of \ac{LRB}. In contrast, the \ac{NN}-aided schemes demand significantly fewer operations: NN-$1$ and NN-$2$ incur approximately $15\%$ and $20\%$ of the Chase-II decoder's FLOPS, respectively. The \ac{BDD} decoder remains the least complex, albeit with the poorest performance.
It is worth noting that these complexity estimates are based on CPU execution. When deployed on hardware platforms with dedicated neural processing capabilities—such as GPUs or specialized accelerators—the cost of the NN-$\rho$ decoders can be further reduced, potentially approaching the complexity of \ac{BDD} while retaining their superior decoding performance. 

In addition to theoretical complexity, Table~\ref{tab:complexity} reports the empirical execution times measured in MATLAB on an Intel Core i7 processor running Ubuntu. The execution time trends are consistent with the \ac{FLOPS} analysis but amplify the gap between Chase-II and the other schemes. This is likely due to implementation inefficiencies in MATLAB's \ac{BDD} function, which may require more operations than predicted analytically. As the Chase-II decoder invokes multiple instances of \ac{BDD}, this inefficiency compounds, resulting in considerably higher runtime.

\begin{table}[t]
   \caption{Complexity analysis of the decoding schemes}
  \label{tab:complexity}
  \centering
  \normalsize
    \begin{tabular}{l|l|l}
    \toprule
    \textbf{Decoder} & \textbf{FLOPS} & \textbf{Runtime (s)}\\
    \midrule
    BDD & $1.5 \cdot 10^3$ & $0.007345$ \\
    Chase-II & $8.19 \cdot 10^6$ & $2.115847$ \\
    NN-$0$ & $1.18 \cdot 10^6$ & $0.012314$ \\
    NN-$1$ & $1.27 \cdot 10^6$ & $0.034333$ \\
    NN-$2$ & $1.63 \cdot 10^6$ & $0.133051$ \\

    \bottomrule
 \end{tabular}
\end{table}

\section{Conclusion} \label{sec:conclusion}
In this paper, we proposed an \ac{NN}-aided decoding framework that significantly reduces the complexity of the Chase decoder while retaining its near-optimal decoding performance. By leveraging a trained \ac{NN} to predict the most promising perturbation pattern, the proposed scheme narrows the set of test patterns that must be evaluated, thereby reducing the number of \ac{BDD} invocations. The resulting decoding process achieves a favorable trade-off between reliability and computational efficiency, making it well suited for \ac{URLLC}.
Extensive simulations on a BCH$(127,64)$ code show that the proposed \ac{NN}-aided decoder achieves a \ac{CER} close to that of the classical Chase decoder, while reducing the required number of floating-point operations and execution time by over $80\%$.
Among the possible future research directions there is the possibility to develop a unified \ac{NN} architecture that generalizes across different \ac{SNR} levels, avoiding the need to train a separate model for each operating point. Moreover, extending the \ac{NN}-aided approach to other classes of quasi-\ac{ML} decoders may unlock further performance and complexity gains in different decoding scenarios.

\section*{Acknowledgments}
This work was partially supported by the European Union under the Italian National Recovery and Resilience Plan (NRRP) of NextGenerationEU, partnership on ``Telecommunications of the Future" (PE00000001 - program ``RESTART").

\bibliographystyle{IEEEtran}
\bibliography{IEEEabrv,ETbib}

\end{document}